\documentclass{PoS}

\newcommand{\be}{\begin{equation}}
\newcommand{\bea}{\begin{eqnarray}}
\newcommand{\ee}{\end{equation}}
\newcommand{\eea}{\end{eqnarray}}

\def\1eq#1{Eq.~(\ref{#1})}
\def\2eqs#1#2{Eqs.~(\ref{#1}) and~(\ref{#2})}
\def\3eqs#1#2#3{Eqs.~(\ref{#1}),~(\ref{#2}) and~(\ref{#3})}
\def\noeq#1{(\ref{#1})} 

\def\G{\Gamma}

\def\diff{\mathrm{d}}

\title{Canonical transformations in gauge theories\\ 
 with non-trivial backgrounds}

\ShortTitle{Canonical transformations and background fields}

\author{Daniele Binosi\\
        European Centre for Theoretical Studies in Nuclear Physics and Related Area (ECT*)\\ and Fondazione Bruno Kessler,\\
 Villa Tambosi, Strada delle Tabarelle 286, I-38123 Villazzano (TN), Italy\\
        E-mail: \email{binosi@ecstar.eu}}

\author{\speaker{Andrea Quadri}\\
        Dept. of Physics, Univ. di Milano and INFN, Sez. di Milano\\
        via Celora 16, I-20133 Milan, Italy\\
        E-mail: \email{andrea.quadri@mi.infn.it}}

\abstract{We show how to implement the background field method
by means of canonical transformations and comment on the applications
of the method to non-perturbative techniques in non-Abelian gauge theories.
We discuss the case of the lattice in some details.}

\FullConference{Sixth International Conference on Quarks and Nuclear Physics,\\
		April 16-20, 2012\\
		Ecole Polytechnique, Palaiseau, Paris}

\begin{document}

\section{Introduction}

Quantization of non-Abelian gauge theories
around topologically non-trivial background fields
plays an important role in understanding their non-perturbative
regime.  Instanton configurations~\cite{Schafer:1996wv} have been
thoroughly investigated since the pioneering work
of 't Hooft~\cite{'tHooft:1976fv}. Another example is provided by
chiral soliton models in effective approaches 
to low-energy QCD~\cite{Meier:1996ng}.

On the other hand, one can also introduce a background gauge
connection as a technical tool for fixing the gauge while retaining 
explicit (background) gauge invariance. 
This leads to the so-called
background field method (BFM) \cite{Abbott:1981ke}.
In the BFM
powerful relations between background and quantum
1-PI amplitudes emerge, which
have been widely used in order to  simplify
computations in many different applications, 
ranging from perturbative Yang-Mills theory
\cite{Ichinose:1981uw}
and the Standard Model \cite{Denner:1994xt} to gravity and supergravity
calculations~\cite{susy}.

The BFM has been formulated on the lattice in~\cite{Luscher:1995vs}, where
the quantization is carried out around a fixed background 
$\hat A_\mu$ with periodic boundary conditions. The approach
of~\cite{Luscher:1995vs} is however purely perturbative, since the path-integral is restricted
to quantum fluctuations in a sufficiently small neighborhood
of $\hat A_\mu$. 

One might then ask whether 
a fully non-perturbative BFM can be implemented on the lattice. In dealing with such a problem,
one has to face several major challenges, and in particular the
well-known Neuberger's 0/0 problem~\cite{Neuberger:1986xz}, which
 prevents the definition
of a consistent non-perturbative BRST symmetry, thus making
unclear how to apply the algebraic procedure, usually adopted for the
perturbative implementation of the BFM~\cite{Grassi:1995wr}.

This difficulty ultimately originates from the existence of Gribov copies:
indeed if one integrates  the BRST-invariant
measure $d\mu_s$  on the gauge fields $A_\mu$, the
auxiliary Nakanishi-Lautrup multiplier  and the ghost and antighost pair 
over a compact cycle (like a 
gauge group orbit in a lattice gauge theory),
one gets zero, due to the fact that on a cycle the gauge-fixing
condition exhibits an even number of solutions, cancelling out pairwise~\cite{Becchi:1996yh}.
It is thus natural
to expect that in a non-perturbative formulation of the BFM there must be
some other mechanism, allowing to control the background dependence
around each of the Gribov solutions.

A possible way of addressing such open issues
is based on the use of canonical transformations
\cite{Binosi:2012pd,Binosi:2012st}. Specifically, it turns out
that whenever the theory is endowed with a Batalin-Vilkovisky (BV) bracket
and satisfies the (extended) Slavnov-Taylor (ST) identity in the presence
of the background field \cite{Binosi:2012pd,Binosi:2011ar}, 
the dependence of the effective
action on the background can be recovered by purely algebraic means
via a canonical transformation ({\it i.e.}, a transformation 
preserving the BV bracket).
In addition, this canonical transformation can be written in an explicit form
by a field-theoretic generalization of the Lie transform,
used in classical analytical mechanics for obtaining finite
canonical transformations once their generating functional is known
\cite{Binosi:2012st}.
 
It is worth stressing that these results are valid
whenever the extended ST identity
holds true. 
In particular, no dynamical ghosts are needed. Therefore, if one is able
to fix the background gauge on the lattice by minimizing some
suitably chosen functional (possibly a natural extension
of the Wilson gauge-fixing functional in the standard lattice
Landau gauge formulation), one might hope to make some progress
toward a non-perturbative formulation of the BFM.
We will comment further on this point 
in Sect.~\ref{sec:app},  after presenting
our main mathematical results 
in Sect.~\ref{sec:res}. 

We refer the reader to ~\cite{Binosi:2012st}
for an overview of the prospects of implementing the BFM  by means of the canonical transformations
in other non-perturbative contexts, {\em e.g.}, in the 2-PI formalism 
or Schwinger-Dyson equations.

\section{Reconstruction of the background dependence via canonical transformations}\label{sec:res}

For a SU(N) Yang-Mills theory, the BV bracket is defined as~\cite{Binosi:2012st}
(only left derivative assumed in what follows)
\bea
\{X,Y\} &=& \int\!\diff^4x \sum_\phi
\left[ (-1)^{\epsilon_{\phi} (\epsilon_X+1)}
\frac{\delta X}{\delta \phi} \frac{\delta Y}{\delta \phi^*}
-(-1)^{\epsilon_{\phi^*} (\epsilon_X+1)}
\frac{\delta X}{\delta \phi^*} \frac{\delta Y}{\delta \phi}
\right].
\label{BVbracket}
\eea
The sum runs over the fields $\phi = (A^a_\mu,c^a)$ and the 
corresponding antifields 
$\phi^* = (A^{*a}_\mu, c^{*a})$, with $\epsilon_\phi$, $\epsilon_{\phi^*}$ and $\epsilon_X$ representing the statistics of the field $\phi$, the antifield $\phi^*$ and the functional $X$ respectively. 

In a linear covariant background gauge, the dependence on the antighost field
$\bar c^a$ can only happen through 
the combination $\widetilde A^{*a}_\mu = A^{*a}_{\mu} + (\widehat {\cal D}_\mu \bar c)^a$, as a consequence of
the antighost equation
$\frac{\delta\Gamma}{\delta {\bar{c}^a}}=-\widehat{\cal D}^{ab}_\mu\frac{\delta\Gamma}{\delta {A^{*b}_\mu}}+({\cal D}^\mu \Omega_\mu)^a$.
$\hat A_\mu$ denotes the background gauge field and $\Omega_\mu$ its
BRST partner~\cite{Grassi:1995wr}. $\widehat {\cal D}_\mu$ 
stands for the covariant derivative
with respect to (w.r.t.) the background.
One can then define the reduced functional $\tilde \G$~\cite{Binosi:2012st} by eliminating the 
Nakanishi-Lautrup-dependent terms (for they only enter at tree-level, 
as a consequence of the usual equation of motion for the Nakanishi-Lautrup multiplier field). 
 Since we will only
deal with the reduced vertex functional in what follows, we will simply
denote it by $\G$.
It obeys the extended ST identity in the presence
of a background  \cite{Binosi:2012pd}:
\be
\int\!\diff^4x\, \Omega^a_\mu(x)
\frac{\delta \G}{\delta \widehat A^a_\mu(x)} = 
- \frac{1}{2}\, \{\G,\G\} .
\label{m.1}
\ee

If one now takes the derivative of \1eq{m.1} w.r.t. $\Omega^a_\mu$ and set the latter source equal to zero afterwards, the resulting equation~\cite{Binosi:2012pd}
\bea
\left.\frac{\delta \G}{\delta \widehat A^a_\mu(x)}\right|_{\Omega=0} = 
\left.- \{ \frac{\delta \G}{\delta \Omega^a_\mu(x)},
\G \}\right|_{\Omega=0},
\label{m.2}
\eea
shows that the 
derivative of the vertex functional w.r.t. 
the background field equals the 
effect of an infinitesimal canonical
transformation (w.r.t. the BV bracket)
on the vertex functional itself. Then,
since the BV bracket does not depend
on either $\widehat A^a_\mu$ or $\Omega^a_\mu$, if one were able to write the finite
canonical transformation generated by the fermion 
$\Psi^{a}_\mu(x)=\frac{\delta \G}{\delta \Omega^a_\mu(x)}$, one would control the full dependence of $\G$
on the background fields; and this would happen not only at the level of the counterterms  of $\G$, but rather for the full 1-PI Green's functions, thus giving control even over the non-local dependence on the background.  

The problem can be thus stated mathematically as follows: given the field and antifield variables $\phi$, $\phi^*$, which are canonical w.r.t. the BV bracket~\noeq{BVbracket}, {\it i.e.},
\bea
\{\phi_i(x),\phi_j(y)\}&=&\{\phi^*_i(x),\phi^*_j(y)\}=0\nonumber \\
\{\phi_i(x),\phi^*_j(y)\}&=&\delta_{ij}\delta^4(y-x),\nonumber
\eea
and the background field $\widehat{A}^a_\mu$,
find the canonical mapping 
$$
(\phi(x),\phi^*(x);\widehat{A}^a_\mu(x))\mapsto(\Phi(x),\Phi^*(x)),
$$
to the new field and antifield variables $\Phi$ and $\Phi^*$
such that the ST identity~\noeq{m.2}, written in these new variables, is automatically satisfied. 
This last condition translates 
into determining the canonical variables $\Phi$ and $\Phi^*$ which are also solutions of the two equations
\bea
\frac{\delta\Phi(y)}{\delta \widehat{A}^a_\mu(x)}&=&\frac{\delta\Psi^a_\mu(x)}{\delta \Phi^*(y)}=\{\Phi(y),\Psi^a_\mu(x)\},\nonumber \\
\frac{\delta\Phi^*(y)}{\delta \widehat{A}^a_\mu(x)}&=&-\frac{\delta\Psi^a_\mu(x)}{\delta \Phi(y)}=\{\Phi^*(y),\Psi^a_\mu(x)\}.
\label{STcond}
\eea

One can obtain the vertex functional $\G$ expressed in terms of the 
canonically transformed variables by means of homotopy 
techniques~\cite{Binosi:2012pd}.  This solution fails to respect the 
(naively expected) exponentiation pattern, due to the dependence of the generating functional $\Psi^a_\mu$ on the background field~$\widehat{A}^a_\mu$.

The explicit canonical transformation
for the fields and antifields of the theory can be 
explicitly worked out as follows. 
We introduce the operator~\cite{Binosi:2012st}
$$
\Delta_{\Psi^{a}_\mu(x)}=\{\cdot,\Psi^{a}_\mu(x)\}+\frac\delta{\delta \widehat{A}^a_\mu(x)} \, , 
$$
where the first term above represents a (graded) generalization (to the BV bracket and a fermionic generator) of the classical Lie derivative w.r.t a (bosonic) generator (in which case the bracket would be the usual Poisson bracket),
while the second term takes into account the above observation on the exponentiation failure.

Next, using the properties of the BV bracket, one can establish the following relations 
\bea
& & \Delta_{\Psi^a_\mu(x)} (\alpha X + \beta Y) =
\alpha \Delta_{\Psi^a_\mu(x)} X + \beta
\Delta_{\Psi^a_\mu(x)} Y, \nonumber \\
& &  \Delta_{\Psi^a_\mu(x)} (XY) = 
X \Delta_{\Psi^a_\mu(x)} Y + (-1)^{\epsilon_X \epsilon_Y} Y \Delta_{\Psi^a_\mu(x)} X,  \nonumber\\
& & \Delta_{\Psi^a_\mu(x)} \{ X, Y \} = 
 \{  \Delta_{\Psi^a_\mu(x)} X, Y \}
+ \{  X,  \Delta_{\Psi^a_\mu(x)} Y \}.\nonumber
\eea
We see then that  $\Delta_{\Psi^a_\mu(x)}$ gives rise to a 
graded derivation with the usual statistics, 
while the last formula allows us to determine the important result 
\bea
\int_1\!\cdots\int_n\,
\widehat A_1 \cdots\hat A_n\,
\Delta_{\Psi_n} \cdots \Delta_{\Psi_1}
\{ X, Y \} &=&
\sum_{0 \leq m \leq n}
\pmatrix{n \cr m} \{
\Delta_{\Psi_1} \cdots \Delta_{\Psi_m} X, 
\Delta_{\Psi_{m+1}}
\dots \Delta_{\Psi_n}Y \},\nonumber\\
\label{20n}
\eea
where we have introduced the shorthand notation \mbox{$\int_i=\int\diff^4y_i$}, \mbox{$\widehat{A}_i=\widehat{A}^{a_i}_{\mu_i}(y_i)$} and $\Psi_i=\Psi^{a_i}_{\mu_i}(y_i)$. 

From the operator $\Delta_\Psi$ one can then define a mapping $E_{\Psi}$, given in terms of a formal power series in the background field $~\widehat{A}$ as follows
\bea
\Phi(x) &=& E_{\Psi}(\phi(x)) 
\label{cantr}\\
&\equiv& \sum_{n \geq 0}
\frac{1}{n!}  \int_1\! \cdots\! \int_n\!
\widehat A_1\cdots\widehat A_n  
\left [ \Delta_{\Psi_n}\! \cdots \Delta_{\Psi_1} \phi(x) \right ]_{\widehat A=0}, \nonumber 
\eea
with an identical expansion holding for the antifields variables. Then~\1eq{cantr} constitutes the sought for canonical mapping between the old and the new variables.  

The canonicity property is a direct consequence of \1eq{20n} above, since the latter directly implies the identity $E_{\Psi}\{X,Y\}=\{E_{\Psi}X,E_{\Psi}Y\}$.
On the other hand, to see that the new variables are indeed solutions of Eqs.~(\ref{STcond}), let us consider first the case of a bosonic field $\Phi$ and expand both the latter and the fermionic generator $\Psi^a_\mu$ in power series w.r.t. the background field $\widehat{A}$. Schematically, one has
\bea
\Phi&=&\phi+\sum_{n \geq 0}
\frac{1}{n!}  \int_1\! \cdots\! \int_n\!
\widehat A_1\cdots\widehat A_n \Phi_{1\cdots n},\nonumber \\
\Psi_0&=&\psi_0+\sum_{n \geq 0}
\frac{1}{n!}  \int_1\! \cdots\! \int_n\!
\widehat A_1\cdots\widehat A_n \Psi_{01\cdots n},\nonumber
\eea
and finds, up to third order in $\widehat{A}$
\bea
& & \Delta_{\Psi_1}\phi\vert_{\widehat{A}=0}=\{\phi,\psi_1\},\nonumber \\
& & \Delta_{\Psi_2}\Delta_{\Psi_1}\phi\vert_{\widehat{A}=0}=\{\{\phi,\psi_1\},\psi_2\}+\{\phi,\Psi_{12}\},\nonumber 
\\
& & \Delta_{\Psi_3}\Delta_{\Psi_2}\Delta_{\Psi_1}\phi\vert_{\widehat{A}=0}=\{\{\{\phi,\psi_1\},\psi_2\},\psi_3\}+\{\phi,\Psi_{123}\}\nonumber 
\\
&& + \{\{\phi,\Psi_{12}\},\psi_3\}+ \{\{\phi,\Psi_{23}\},\psi_1\}+ \{\{\phi,\Psi_{31}\},\psi_2\},\nonumber 
\eea
where in the last equation we have symmetrized all indices, and used the (graded) Jacobi identity together with the result $
\int_1\!\int_2\!\widehat{A}_1\widehat{A}_2\frac{\delta}{\delta\widehat{A}_3}\{\Psi_1,\Psi_2\}=0.
$
It can then be  checked that the above terms are indeed the solutions (up to third order in $\widehat{A}$) of the first of Eqs.~(\ref{STcond}). The fermionic case, {\it e.g.},  a fermionic antifield $\Phi^*$, can be treated in the same way.

\medskip
Let us end by observing that
there is a close relation with the theory of Lie transforms
in classical analytical mechanics. Suppose that one wants
to find out the finite canonical transformation, depending
on the small parameter $\epsilon$, which maps the canonically
conjugated variables $(q,p)$ into the new ones $(Q,P)$, under
the assumption that $Q,P$ satisfy the following differential equations:
$$\frac{\diff Q}{\diff \epsilon} = \frac{\partial}{\partial P} V(p,q;\epsilon); \,  \qquad \frac{\diff P}{\diff \epsilon} = - \frac{\partial}{\partial Q} V(p,q;\epsilon) \, .$$
The generating functional $V(p,q;\epsilon)$ of the canonical transformations
depends on a small parameter $\epsilon$.  Then the solution is obtained
by a Lie series of the operator $\Delta_{V} = \{ \cdot, V\} + 
\frac{\partial}{\partial \epsilon}$, where the bracket is here
the usual Poisson bracket~\cite{Binosi:2012st}.

\section{Lattice background field method}\label{sec:app}

On the lattice, the background gauge-fixing can be implemented
by minimizing the functional~\cite{Binosi:2012st,Cucchieri:2012ii}
\bea
F[g] = - \int \diff^4 x ~ {\rm Tr} (A_\mu^g - \hat A_\mu)^2
\eea
w.r.t. the group element $g$.
$A_\mu^g$ is the gauge transform of the gauge field $A_\mu$. 
After minimizing $F[g]$, one gets the background Landau gauge condition
$\widehat {\cal D}_\mu (A_\mu^g - \hat A_\mu)=0$.
The canonical transformation, reconstructing the dependence
on the background field, is the one induced by
the mapping $A'_\mu =A_\mu^{g(A,\hat A)} - \hat A_\mu$, where the new
gauge field $A'_\mu$ obeys the Landau gauge condition.

If, as a consequence of the
existence of the Gribov copies, multiple minima are present,
they
will be parameterized by different functions $g_i(A,\hat A)$.
Each of them controls in principle the dependence on the background, provided
that one restrict himself to the region of validity of each~$g_i$.

\section{Conclusions}

We have reviewed the approach to the implementation of the BFM
based on  canonical transformations
proposed in \cite{Binosi:2012st}. This approach has a very
general range of applicability: it encompasses the
standard perturbative treatment of the BFM and provides
a way to extend the BFM in a systematic way to non-perturbative
settings. In the lattice case, it can in principle 
disentangle the solutions
related to different Gribov copies, thus allowing to 
obtain the canonical mapping which governs 
the dependence on the background field around each of the minima
of the gauge-fixing functional.
In addition, notice that
the comparison between the continuum and the lattice
computations might be greatly eased by the powerful relations between 1-PI 
background and quantum amplitudes, stemming
from the background gauge invariance.

Finally, our approach could also be applied to non-perturbative
frameworks in the continuum, like the 2-PI formalism and the
Schwinger-Dyson equations.


\begin{thebibliography}{99}

\bibitem{Schafer:1996wv}
  For a review see {\it e.g.} T.~Schafer and E.~V.~Shuryak,
  Rev.\ Mod.\ Phys.\  {\bf 70} (1998) 323.

\bibitem{'tHooft:1976fv}
  G.~'t Hooft,
  Phys.\ Rev.\ D {\bf 14} (1976) 3432
   [Erratum-ibid.\ D {\bf 18} (1978) 2199].

\bibitem{Meier:1996ng}
  F.~Meier and H.~Walliser,
  Phys.\ Rept.\  {\bf 289} (1997) 383.

\bibitem{Abbott:1981ke}
 L.~F.~Abbott,
  Acta\ Phys.\ Polon.\ {\bf B13} (1982) 33;
%
 Nucl.\ Phys.\ B {\bf 185} (1981) 189.

\bibitem{Ichinose:1981uw}
  S.~Ichinose and M.~Omote,
  Nucl.\ Phys.\ B {\bf 203} (1982) 221;
  D.~M.~Capper and A.~MacLean,
  Nucl.\ Phys.\ B {\bf 203} (1982) 413.

\bibitem{Denner:1994xt}
  A.~Denner, G.~Weiglein and S.~Dittmaier,
  Nucl.\ Phys.\ B {\bf 440} (1995) 95;
  P.~A.~Grassi, T.~Hurth and M.~Steinhauser,
  Nucl.\ Phys.\ B {\bf 610} (2001) 215.

\bibitem{susy}
S.~J.~Gates, M.~T.~Grisaru, M.~Rocek et al., Front.\ Phys.\ {\bf 58} (1983) 1.

\bibitem{Luscher:1995vs}
  M.~Luscher and P.~Weisz,
  Nucl.\ Phys.\ B {\bf 452} (1995) 213.

\bibitem{Neuberger:1986xz}
  H.~Neuberger,
  Phys.\ Lett.\ B {\bf 183} (1987) 337.

\bibitem{Grassi:1995wr}
  P.~A.~Grassi,
  Nucl.\ Phys.\ B {\bf 462} (1996) 524;
  C.~Becchi and R.~Collina,
  Nucl.\ Phys.\ B {\bf 562} (1999) 412; 
  R.~Ferrari, M.~Picariello and A.~Quadri,
  Annals Phys.\  {\bf 294} (2001) 165.

\bibitem{Becchi:1996yh}
  C.~Becchi,
  ``Introduction to BRS symmetry,''
  hep-th/9607181.

\bibitem{Binosi:2012pd}
  D.~Binosi and A.~Quadri,
  Phys.\ Rev.\ D {\bf 85} (2012) 085020. 

\bibitem{Binosi:2012st}
  D.~Binosi and A.~Quadri,
  ``The Background Field Method as a Canonical Transformation,''
  arXiv:1203.6637 [hep-th]. To appear in Phys.\ Rev.\ D - Rapid communications.

\bibitem{Binosi:2011ar}
  D.~Binosi and A.~Quadri,
  Phys.\ Rev.\ D {\bf 84} (2011) 065017.

\bibitem{Cucchieri:2012ii}
  A.~Cucchieri and T.~Mendes,
  arXiv:1204.0216 [hep-lat].

\end{thebibliography}
\end{document}